\documentclass[10pt,letterpaper]{article}
\usepackage{opex3}

\usepackage{graphicx}
\usepackage{amsmath,amssymb}
\usepackage{dcolumn}
\usepackage{multirow}
\usepackage{bm}
\usepackage{cite}

\begin{document}

\title{Field-test of a robust, portable, frequency-stable laser}

\author{David R. Leibrandt,* Michael J. Thorpe, James C. Bergquist, \\ and Till Rosenband}
\address{National Institute of Standards and Technology, 325 Broadway St., Boulder, CO 80305, USA}
\email{*david.leibrandt@nist.gov}

\date{\today}

\begin{abstract*}
We operate a frequency-stable laser in a non-laboratory environment where the test platform is a passenger vehicle.  We measure the acceleration experienced by the laser and actively correct for it to achieve a system acceleration sensitivity of $\Delta f / f$ = $11(2) \times 10^{-12}$/g, $6(2) \times 10^{-12}$/g, and $4(1) \times 10^{-12}$/g for accelerations in three orthogonal directions at 1 Hz.  The acceleration spectrum and laser performance are evaluated with the vehicle both stationary and moving.  The laser linewidth in the stationary vehicle with engine idling is 1.7(1) Hz.
\\
\end{abstract*}


\bibliographystyle{osajnl}

\section{Introduction}

Frequency-stable lasers \cite{Young1999subHzLaser} are important tools whose present applications include optical frequency standards \cite{Chou2009AlAl}, gravitational wave detection \cite{Willke2008}, and tests of fundamental physics \cite{TR2008AlHg,Blatt2008SrGravity}.  Stability is most often achieved by locking the laser to a Fabry-P\'{e}rot cavity.  This technique transfers the length stability of the cavity to frequency stability of the laser.  While significant progress towards miniaturization \cite{Ludlow2007Football} and transportability \cite{Vogt2010TransportableLaser} has been achieved, thus far the best performing lasers have been constrained to operate in well controlled laboratory environments due primarily to reference cavity acceleration sensitivities greater than about $10^{-10}$/g \cite{Webster2007VibrationInsensitive,Millo2009VibrationInsensitive}.  There is growing interest in frequency-stable lasers capable of operation outside the laboratory for applications such as geodesy \cite{Kleppner2006Geodesy,Chou2010Relativity}, hydrology \cite{Schmidt2008GRACE}, and space-based tests of fundamental physics \cite{Schiller2009SpaceClocks,Wolf2009SpaceClocks}.

In this work, we operate a cavity-stabilized laser in a non-laboratory environment.  A passenger vehicle serves as the test platform.  Switching the vehicle engine on allows us to test the laser performance in a vibrationally noisy environment, and driving the vehicle allows us to measure the laser's frequency response to transport dynamics.  Low passive acceleration and orientation sensitivity of the reference cavity are achieved by a design in which a spherical cavity spacer is held rigidly at two points on a diameter of the sphere \cite{Leibrandt2010SC}.  The support axis orientation of 53$^\circ$ with respect to the optical axis reduces coupling of acceleration-induced squeeze forces to the cavity length.  Acceleration sensitivity is further reduced by measuring the acceleration of the cavity and applying real-time feed-forward corrections to the laser frequency \cite{Thorpe2010FeedForward}.  We test the laser both when the vehicle is stationary and when it is moving with peak accelerations greater than 0.1~g.

This paper proceeds as follows: Section \ref{sec:setup} describes the test setup, including the real-time acceleration feed-forward scheme.  Section \ref{sec:stationary} discusses the acceleration spectrum and laser performance when the vehicle is stationary as well as the acceleration sensitivity of the laser.  Section \ref{sec:mobile} discusses the laser performance when the vehicle is moving.  Finally, Section \ref{sec:conclusion} summarizes and concludes.

\section{Test setup}\label{sec:setup}

\begin{figure}
\begin{center}
\includegraphics[width=0.8\columnwidth]{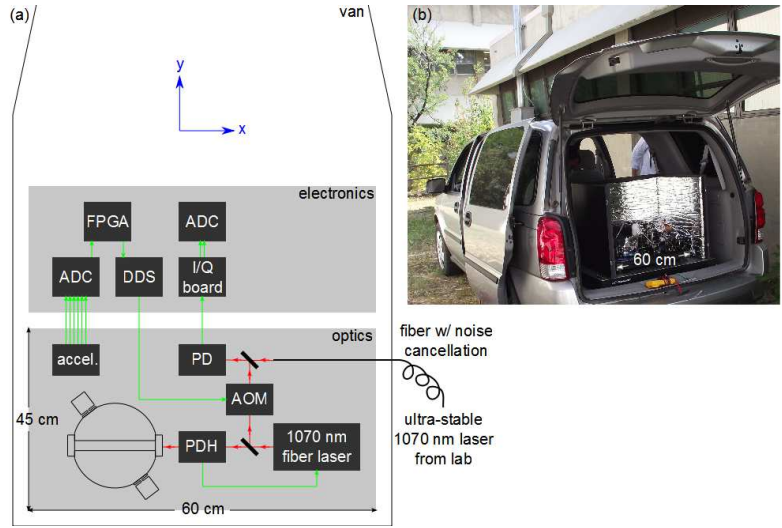}
\caption{\label{fig:setup}Experimental setup.  (a) Schematic of the test platform.  A cavity-stabilized laser and the associated electronics are housed in a passenger vehicle in order to evaluate the performance of the laser in a field environment.  The phase of the test laser is measured with respect to the phase of a laboratory-based reference laser via a fiber-optic link.  (b) Photograph of the test platform showing the optical breadboard inside the thermal enclosure (open) in the back of the vehicle.  FPGA: field-programmable gate array, ADC: analog-to-digital converter, DDS: direct-digital synthesizer, I/Q: in-phase/quadrature, PD: photodiode, AOM: acousto-optic modulator, PDH: Pound-Drever-Hall detector.}
\end{center}
\end{figure}

The setup consists of an optical breadboard, electronics for locking the laser and applying the acceleration feed-forward, and a computer for data acquisition, all of which are housed in a passenger vehicle \cite{ChevyUplander,CommercialProducts} (see Figure \ref{fig:setup}).  The vehicle is tethered to a building by a power cable and an optical fiber that transmits light from a reference laser located inside our laboratory for measurements of the test laser frequency.  This limits the driving range to about 5 m over a grass surface.

The optical breadboard holds the test laser, the cavity, the optics for locking the laser to the cavity, and the optics for making frequency comparisons with the reference laser.  The breadboard is an aluminum honeycomb construction with dimensions of 60~cm by 45~cm by 5~cm.  Including the optics, the height is 30~cm and the mass is 34~kg.  The breadboard rests on four viscoelastic polymer cylinders (38~mm diameter by 25~mm tall) for high-frequency vibration isolation.  We do not use active vibration isolation because the majority of accelerations experienced in a moving vehicle have a displacement amplitude that is too large for an active vibration isolation system to correct, i.e., the vehicle dynamics are fundamentally non-inertial.  The cavity is identical to that described in Reference \cite{Leibrandt2010SC}, with a 50.8 mm diameter spherical Corning Ultra Low Expansion (ULE) glass spacer \cite{CommercialProducts} and fused silica mirrors, except that we have added ULE rings \cite{Legero2010Rings} to the backs of the mirrors to raise the zero-crossing temperature of the cavity coefficient of thermal expansion (CTE) from about -25~$^\circ$C to about 7~$^\circ$C.  The cavity temperature is stabilized at the CTE zero-crossing temperature by a thermoelectric cooler (TEC) inside the vacuum chamber, and the entire optical breadboard is enclosed in a box that is temperature stabilized to 30~$^\circ$C by a resistive heater.  Note that the field-test described in this paper relies on a compressive cavity mount to survive the lateral accelerations experienced while driving, and that the mounted cavity is insensitive to changes of the compressive force \cite{Leibrandt2010SC}.  A 1070~nm fiber laser (test laser) is locked to the cavity by the Pound-Drever-Hall (PDH) method \cite{Drever1983PDH} via fast feedback to an acousto-optic modulator (AOM) and slow feedback to a piezo-electric transducer in the laser cavity.  The phase of the test laser is measured via a heterodyne beat note with an independent reference laser that is located inside our laboratory.  This reference laser typically serves as the clock laser that drives the $^1$S$_0$ to $^3$P$_0$ transition in $^{27}$Al$^+$ \cite{Chou2009AlAl}.  The reference laser light arrives through a polarization-maintaining noise-canceled fiber \cite{Ma1994FiberNoise} and is superimposed with the test laser light on a fast photodiode.  The resulting beat note is first mixed down to about 6~MHz, then further mixed to near 0 Hz by a demodulator circuit that outputs the in-phase (I) and quadrature (Q) components of the signal.  The I and Q components are recorded with an analog-to-digital converter (ADC) at a sample rate of either 10~kHz or 100~kHz.

Acceleration feed-forward is implemented via a Wiener filter technique similar to that described in Reference \cite{Thorpe2010FeedForward}.  Instead of correcting the laser phase, however, here we correct the laser frequency.  This allows us to use a Wiener filter with a much shorter time memory.  The best performance is obtained with only one filter coefficient (per accelerometer), which is equivalent to correcting for an acceleration transfer function that is constant in frequency over the bandwidth of the accelerometer.  This configuration minimizes the latency between measuring the acceleration and applying the feed-forward correction.  The filter coefficients are derived from simultaneous measurements of the cavity accelerations and laser frequency noise with the vehicle engine on and the vehicle stationary for 100~s.  All of the measurements presented in this paper used the same filter coefficients, and we did not observe degradation in feed-forward performance over a period of two weeks.  Accelerations are measured by six one-axis, piezo-electric accelerometers \cite{PCBaccel,CommercialProducts} mounted around the cavity in a geometry where there is one sensor at the center of each face of a cube, with the sensing axis oriented on a diagonal of the cube face such that these diagonals form a regular tetrahedron \cite{Chen1994AccIMU}.  The center of the cube is located near the center of the cavity.  This arrangement of sensors allows us to measure both the linear and rotational acceleration in the limit that the centripetal acceleration is small.  The accelerometers have a bandwidth of 0.06~Hz to 450~Hz and their signals are input into a field-programmable gate array (FPGA) that computes the predicted laser frequency by use of the stored Wiener filter.  After the predicted frequency is calculated, the FPGA corrects the laser frequency by changing the output frequency of a direct digital synthesizer (DDS) that drives an AOM.  The feed-forward update rate is approximately 10 kHz, and the latency between measuring the acceleration and applying the feed-forward correction is less than 100~$\mu$s.

\section{Stationary laser stability and acceleration sensitivity}\label{sec:stationary}

\begin{figure}
\begin{center}
\includegraphics[width=0.6\columnwidth]{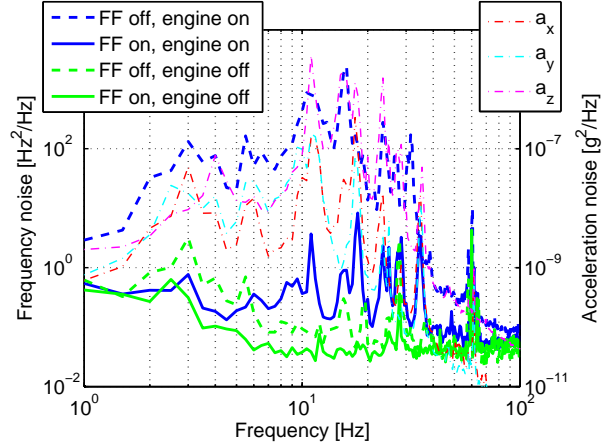}
\caption{\label{fig:frequencyNoise}Laser frequency noise and acceleration power spectral densities with the vehicle stationary.  The laser frequency noise is shown for four cases, with the acceleration feed-forward (FF) and the vehicle engine on and off.  Acceleration noise is shown with the engine on and corresponds to an RMS value of 3.3~mg.  With the engine on, the acceleration feed-forward suppresses the laser frequency noise by over 20~dB between 3 and 30~Hz.  The laser frequency noise is lower yet with the vehicle engine off, which shows that the acceleration feed-forward is not able to remove all of the acceleration-based laser frequency noise.}
\end{center}
\end{figure}

\begin{figure}
\begin{center}
\includegraphics[width=0.6\columnwidth]{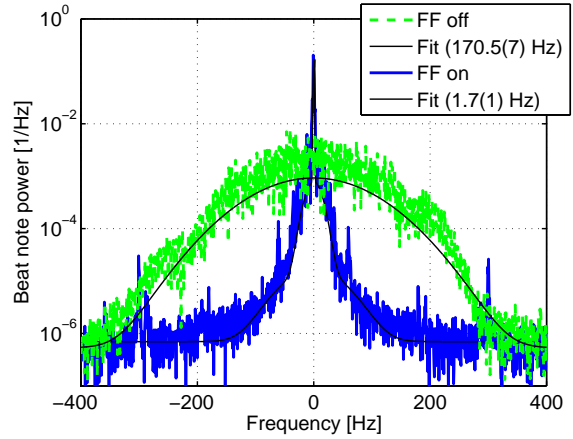}
\caption{\label{fig:beatNote}Power spectrum of the beat note with the vehicle engine on and the vehicle stationary for 5~s of data (resolution bandwidth = 0.6~Hz). The spectrum is shown both with the acceleration feed-forward (FF) off (170.5(7)~Hz FWHM) and the acceleration feed-forward on.  With the acceleration feed-forward on there is a 1.7(1)~Hz FWHM coherent peak on top of a pedestal, with 77~\% of the power in the coherent peak.  In both cases the beat note is normalized such that the total power is 1.}
\end{center}
\end{figure}

The first test was performed with the vehicle stationary and the engine either off or on (idle).  Figure \ref{fig:frequencyNoise} shows the laser frequency and acceleration power spectral densities.  The acceleration noise is shown with the vehicle engine on.  It has a broad peak between 10~Hz and 20~Hz corresponding to the rotational frequency of the engine and a root-mean-square (RMS) value of 3.3~mg (1~g = 9.8~m/s$^2$).  When the acceleration feed-forward is turned on, the laser frequency noise is suppressed by greater than 20~dB between 3~Hz and 30~Hz where most of the acceleration noise resides.  The remaining acceleration-based laser frequency noise may be due to the angular velocity of the cavity, which we do not measure.   We attempted to obtain the angular velocity by numerical integration of the angular acceleration and incorporate it into the Wiener filter as well, but the integral diverged too quickly to be useful.  Figure \ref{fig:beatNote} shows the power spectrum of the beat note between the test laser and the reference laser with the vehicle engine on and acceleration feed-forward both off and on.  We fit the beat note with the acceleration feed-forward on to a 1.7(1)~Hz full-width at half-maximum (FWHM) coherent peak above a pedestal, with 77~\% of the power in the coherent peak.

\begin{figure}
\begin{center}
\includegraphics[width=0.6\columnwidth]{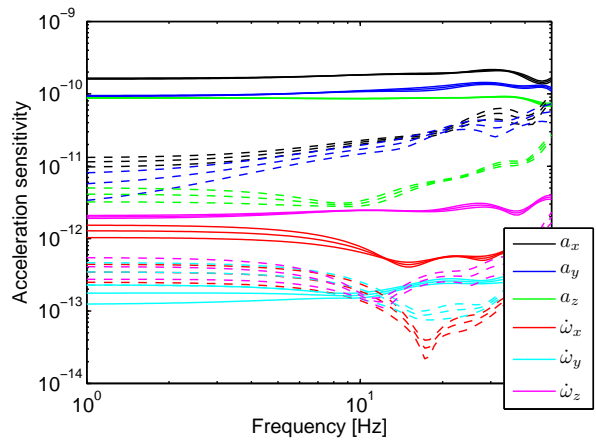}
\caption{\label{fig:accelerationSensitivity}Acceleration sensitivity of the reference cavity (solid lines) and the actively corrected reference cavity (dashed lines) as a function of frequency.  Linear acceleration sensitivities are plotted in units of $(\Delta f / f) / g$ and angular acceleration sensitivities are plotted in units of $(\Delta f / f) / (\textrm{rad/s}^2)$.  The three lines of each type are the mean (center curve) and standard error (lower and upper curves).}
\end{center}
\end{figure}

Figure \ref{fig:accelerationSensitivity} shows the linear and angular acceleration sensitivity of the reference cavity alone and of the combined reference cavity with acceleration feed-forward system.  These sensitivities were obtained by simultaneously measuring the acceleration and the laser frequency with the vehicle engine on and the vehicle stationary for 100 s, computing the Wiener filter for 500 filter coefficients, and taking the discrete Laplace transform of the filter coefficients to obtain the acceleration transfer functions.  The acceleration sensitivity of the actively corrected reference cavity in the two horizontal and the vertical directions at 1 Hz is $11(2) \times 10^{-12}$/g, $6(2) \times 10^{-12}$/g, and $4(1) \times 10^{-12}$/g.  At higher frequency, the system acceleration sensitivity is signal-to-noise ratio limited because there is less acceleration noise to correct.

\section{Mobile laser stability}\label{sec:mobile}

\begin{figure}
\begin{center}
\includegraphics[width=0.6\columnwidth]{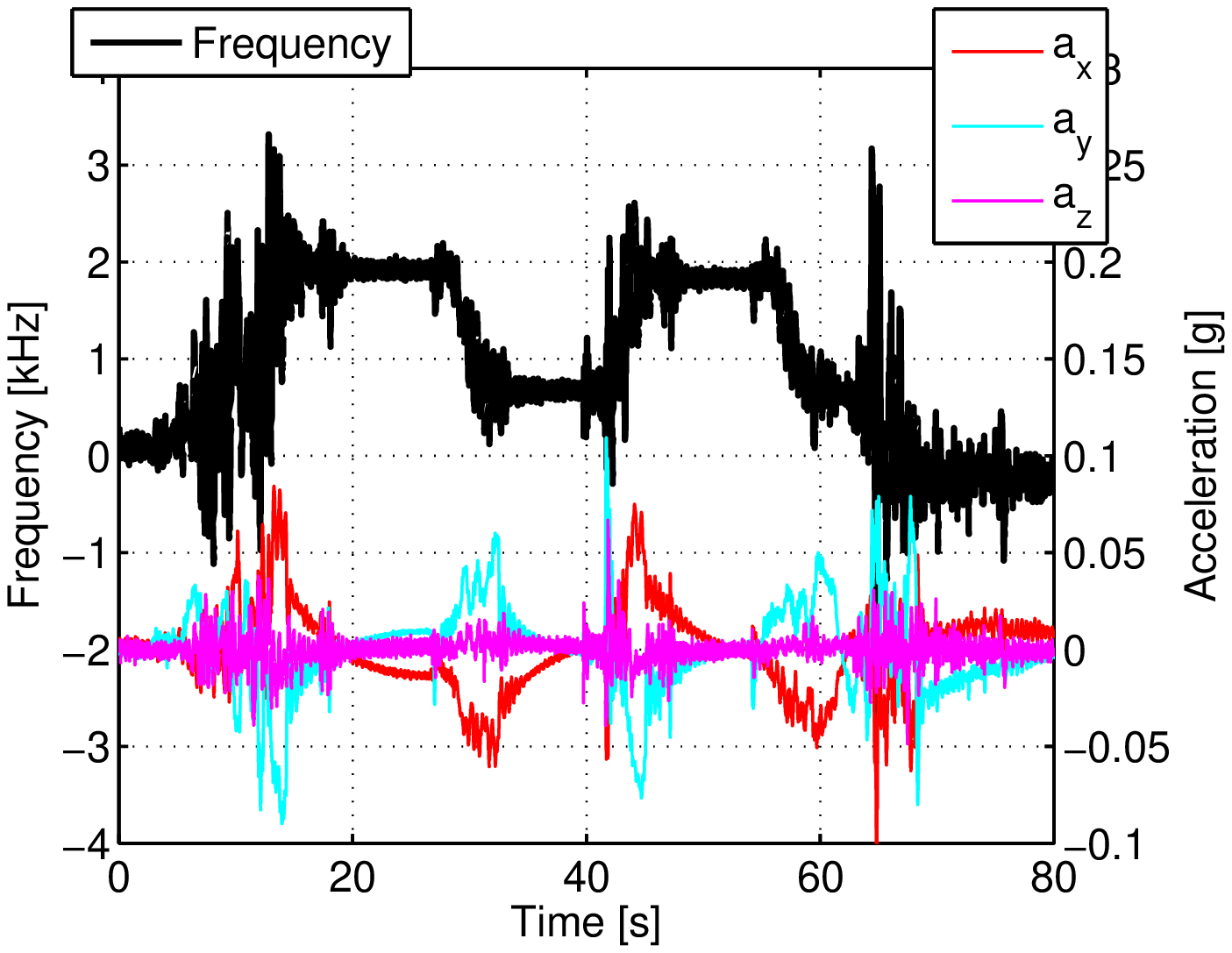}
\includegraphics[width=0.6\columnwidth]{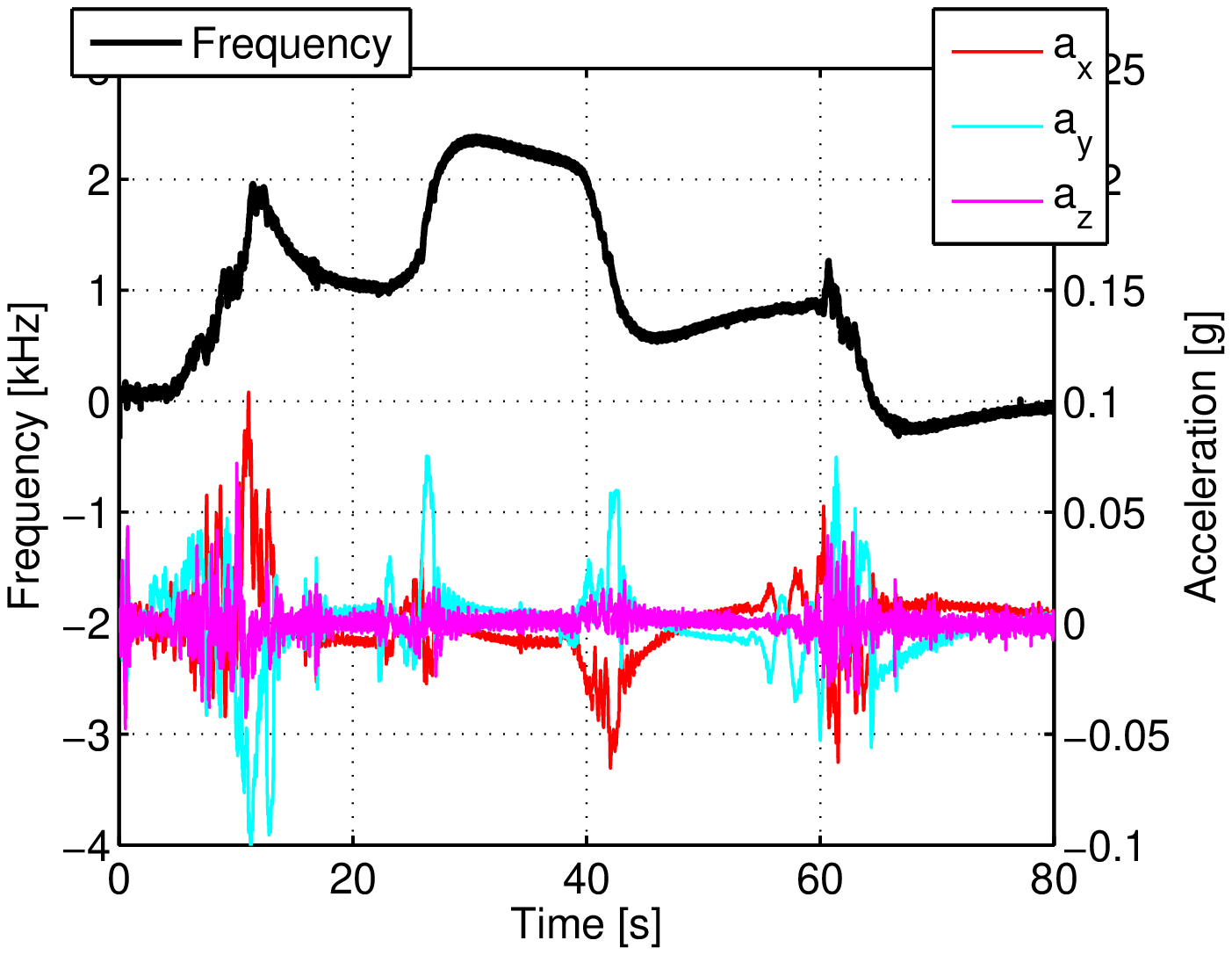}
\caption{\label{fig:freqVsTimeDriving}Laser frequency and acceleration as a function of time while driving with the acceleration feed-forward off (top) and on (bottom).  The acceleration feed-forward suppresses the high-frequency laser frequency noise due to driving over bumps but is unable to correct for frequency shifts due to changes in orientation of the reference cavity with respect to gravity.}
\end{center}
\end{figure}

We performed a second set of measurements while driving over an uneven surface.  Vehicle motion was limited to low speeds (less than 1 m/s) because at higher speeds the reference laser fiber noise cancellation was unreliable.  However, even at low speeds the accelerations were as large as 0.1 g.  Figure \ref{fig:freqVsTimeDriving} shows the laser frequency and acceleration for two measurements with the acceleration feed-forward off and on.  For each measurement we started and stopped driving several times.  With the acceleration feed-forward off, there is approximately 1 kHz amplitude high-frequency noise due to driving over bumps and up to 2 kHz frequency shifts due to changes in orientation of the reference cavity with respect to gravity.  The measured passive acceleration sensitivity of our reference cavity implies that a 2 kHz shift corresponds to a $4^\circ$ change in orientation of the reference cavity, consistent with the uneven driving surface.  With the acceleration feed-forward on, the high-frequency noise is suppressed but the frequency shifts are still present.  This is because our piezo-electric accelerometers have no DC response, so at low frequency (below 0.06 Hz) the acceleration sensitivity of the actively-corrected reference cavity is exactly the same as that of the reference cavity alone.

\section{Conclusion}\label{sec:conclusion}

We have demonstrated operation of a frequency-stable laser in a field environment.  In a stationary vehicle with the engine on, we achieve a 1.7(1) Hz FWHM laser in a 3.3 mg RMS acceleration environment.  In a moving vehicle we show approximately 100 Hz FWHM operation with accelerations as large as 0.1 g for time scales less than 1 s.  On longer time scales, we observe up to 2 kHz frequency shifts due to changes in orientation of the reference cavity with respect to gravity that cannot be corrected by the current acceleration feed-forward system.  Orientation changes could be corrected for in future work by use of variable-capacitance MEMS (microelectromechanical systems) accelerometers in place of the piezo-electric accelerometers used here.

The acceleration sensitivity achieved in this work, $11(2) \times 10^{-12}$/g, $6(2) \times 10^{-12}$/g, and $4(1) \times 10^{-12}$/g for accelerations in three orthogonal directions at 1 Hz, is roughly one order of magnitude lower than that of the best radio-frequency crystal oscillators \cite{Hati2007CrystalAccSen}.  To our knowledge, smaller acceleration sensitivies are achieved only in atomic frequency standards \cite{Hellwig1990}.  This laser could be used for acceleration insensitive synthesis of low-noise radio-frequency or microwave signals via a femtosecond frequency comb \cite{Zhang2010comb,Baumann2009vibrationComb}.  The technologies demonstrated in this work, acceleration and orientation insensitive reference cavities and acceleration feed-forward, pave the way for new applications of frequency-stable lasers outside the laboratory.

\section*{Acknowledgments}

We acknowledge contributions to the construction of the cavity from R.~E.~Drullinger, M.~Notcutt, and G.~Hua.  We thank W.~Swann and D.~Howe for critical readings of this manuscript.  This work is supported by ONR, AFOSR, and DARPA.  D. R. Leibrandt and M. J. Thorpe acknowledge support from the National Research Council.  This work is not subject to U.S. copyright.


\begin{thebibliography}{10}
\bibitem{CommercialProducts}
Commercial products are identified for technical clarity. Such identification
  does not imply endorsement by the National Institute of Standards and
  Technology.
\end{thebibliography}
\end{document}